\documentclass[conference]{IEEEtran}
\IEEEoverridecommandlockouts
\usepackage[numbers,sort&compress]{natbib}
\usepackage{amsmath,amssymb,amsfonts}
\usepackage{algorithmic}
\usepackage{graphicx}
\usepackage{textcomp}
\usepackage{xcolor}
\usepackage{verbatim}
\usepackage{bbm}
\usepackage[utf8]{inputenc}
\usepackage[utf8]{inputenc}
\usepackage{amssymb}
\DeclareUnicodeCharacter{211D}{\mathbb{R}}

\usepackage{bbold}
\def\BibTeX{{\rm B\kern-.05em{\sc i\kern-.025em b}\kern-.08em
    T\kern-.1667em\lower.7ex\hbox{E}\kern-.125emX}}

\begin{document}

\title{Comparing Two Different Approaches in Big Data and Business Analysis for Churn Prediction with the Focus on How Apache Spark Employed\\
}

\author{
 \IEEEauthorblockN{Mohammad Sina Kiarostami}
 \IEEEauthorblockA{\textit{Department of Computer Science and Engineering} \\
 \textit{University of Oulu}\\
 Oulu, Finland \\
 mohammad.kiarostami@oulu.fi}

}

\maketitle

\begin{abstract}

Due to the significant importance of Big Data analysis, especially in business-related topics such as improving services, finding potential customers, and selecting practical approaches to manage income and expenses, many companies attempt to collaborate with scientists to find how, why, and what they should analysis.
\par In this work, we would like to compare and discuss two different approaches that employed in business analysis topic in Big Data with more consideration on how they utilized Spark. Both studies have investigated Churn Prediction as their case study for their proposed approaches since it is an essential topic in business analysis for companies to recognize a customer intends to leave or stop using their services. Here, we focus on Apache Spark since it has provided several solutions to handle a massive amount of data in recent years efficiently. This feature in Spark makes it one of the most robust candidate tools to upfront with a Big Data problem, particularly time and resource are concerns.

\end{abstract}


\section{Introduction}

These days, Big Data and business analysis are becoming more alluring in both academic and industrial aspects \cite{iqbal2020big}. In academic areas, researchers try to find more accurate and efficient solutions such as improving algorithms in Big Data or adjusting the architecture of current approaches since resources and time are expensive \cite{agrawal2020technologies}. In the other area, many companies are looking for more accurate and faster approaches due to the significance of truths that could be revealed by analyzing data, such as how to improve the service effectively \cite{ajah2019big}.

Churn prediction \cite{keramati2016developing} is one the hottest topics in the business analysis due to the importance of competitions among companies to not only acquire more customers but also not losing customers since it costs the companies to lose who gets and uses their services \cite{sayed2018predicting, lazarov2007churn}. Churn prediction means predicting and recognizing customers who intend to leave or stop using a company's service or product \cite{sayed2018predicting}. If a customer leaves a service, it would cost that company to lose a customer and for losing more customers due to the avalanche effect. It also can help the companies to improve their services. Therefore, there are many approaches to predict and avoid this issue \cite{huang2012customer}.

Apache Spark is always a well-deserved tool (open-source engine) to let scientists struggle with a huge amount of data when time and resource usage are concerned. It provides many distinct solutions for different issues in Big Data analysis, such as \textit{MapReduce} limitations to get more accurate results in a lesser time \cite{tianren2018modeling}, which is vital for everyone \cite{kiarostami2019multi}. This feature makes Apache Spark a robust and trusted tool in comparison with others such as Apache Hadoop.

In the rest of the paper, in section II, we compare two different approaches \cite{sayed2018predicting, zdravevski2020big} in business analysis with the churn prediction case study by focusing on how they employed Apache Spark. Then, we discuss the approaches and conclude the debate in section II. 

\section{Comparison}

In \cite{sayed2018predicting}, Sayed et al. performed a comparative study to investigate and compare the impacts of two different Apache Spark packages, namely ML and MLlib, regarding accuracy and performance in model training and evaluation. They employed a bank customers transactions dataset to study customer churn prediction. After the comparison, they indicated MLlib package with RDD-based API performs better in training time. However, the ML package with its DataFrames-based API is better regarding testing performance and overall accuracy. So, they selected the ML package for customer churn prediction. 

In \cite{zdravevski2020big}, Zdravevski et al. designed and explained a cloud-based architecture to improve Extract-Transform-Load (ETL) in Big Data and then evaluated their model on a churn prediction study as well in three modifications (Session-based, Predefined time period, and No Aggregation). They employed Spark in the data extraction and transformation phases of their proposed approach. After that, they explained that the results of these phases loaded into a data warehouse while it benefited from edge computing utilization to reduce the workload on the main servers, which are Hadoop clusters on Amazon AWS. They indicated that their proposed solution has several advantages, such as combining Big Data technologies and traditional data warehousing technologies and increasing the capability to upfront with massive datasets (Big Data stored on OSS, 2.6 TB). 

These approaches have a fundamental difference. In \cite{zdravevski2020big}, they propose a complete architecture that Spark is part of it. However, in \cite{sayed2018predicting}, the authors compare two existing solutions in Spark and let say they build up their study based on it. Spark plays the primary role in both approaches with no argument since both studies utilized Spark to process large amounts of data. In \cite{zdravevski2020big}, the approach processes created data lakes with Spark. It also employs Spark clusters to implement complex ETL and Machine Learning algorithms and perform complicated aggregations. ETL processes are reasonably similar to MapReduce processes \cite{bala2014p}. After considering this similarity and the fact that they employed Spark to perform the ETL process, their approach could be comparable from this point of view with \cite{sayed2018predicting} since they used Spark to solve MapReduce limitations.

In \cite{zdravevski2020big}, they also stated that the results of Spark concluded to HDFS or stored on OSS based on the circumstance and requirements, while in \cite{sayed2018predicting}, they utilized this engine to build two models, train, and evaluate them. Although both approaches utilized Spark to process the data in a general view, in \cite{zdravevski2020big}, Spark can help them in using resources dynamically since their solution is designed based on a cluster. Because of Spark's feature, it does not need to reset a task when it recognizes another resource during the execution. In \cite{zdravevski2020big}, Spark not only processes the data but also plays a task manager tool since it distributes the loadings on the available nodes. They also utilized TextFiles and WholeTextFiles operations in Spark to read the new data from OSS, MapPartitions operation to transform data correctly, and finally speed up these steps. At the same time, in \cite{sayed2018predicting}, they employed two packages with APIs in which the data process and improving performance goals are the same as \cite{zdravevski2020big}.

\section{Discussion and Conclusion}

In this work, we investigated and compared two approaches in business analysis and Big Data that both ways have studied customer churn prediction as to their case study. We mainly focused on how these approaches employed Spark due to the significance this open-source engine has in Big Data analysis. In \cite{sayed2018predicting}, the authors used Spark to develop two models based on MLlib and ML packages to recognize which package is more efficient. They showed that due to the internal transformations in both packages, MLlib is better only in training time, and ML outperforms the other in a lesser time with higher accuracy. In \cite{zdravevski2020big}, they employed Spark as a part of their solution to not only extract and transform a considerable amount of data in three scenarios but also store and read data to and from databases and manage the resources in the cluster due to the features of Spark. These approaches utilized Spark as an engine to process their data, and in \cite{zdravevski2020big}, we saw more than processing data from Spark. Both methods showed how and why Spark is a crucial tool in Data analysis and Big Data. 

Finally, Apache Spark is an open-source engine with many operations and features to help scientists improve their research, implement complex algorithms on more significant amounts of data, and increase the capability and performance of their solutions. 

\section{Acknowledgement}

We would like to send our most tremendous appreciation to Dr Lauri Loven for suggesting this work. 

\bibliographystyle{unsrt}
\bibliography{references}

\end{document}